\documentclass[final,3p,times,twocolumn]{elsarticle}
\usepackage{graphicx}
\usepackage{hyperref}

\usepackage{amssymb}
\usepackage{amsmath}
\usepackage{amsthm}
\usepackage{lipsum}
\journal{Nuclear Instruments and Methods in Physics Research Section A}
\begin{document}

\begin{frontmatter}


 %%\author{Ish Mukul\corref{cor1}\fnref{label2}}
%% \ead{ishmukul@gmail.com}

\title{A new scheme to measure the electron-neutrino correlation - the case of $^{6}$He}

\author[a]{I. Mukul}
 \ead{ishmukul@gmail.com}
\author[a]{M. Hass}
\ead{michael.hass@weizmann.ac.il}
\author[a]{O. Heber}
\author[b]{T. Y. Hirsh}
\author[a,b,c]{Y. Mishnayot}
\author[a]{M. L. Rappaport}
\author[c]{G. Ron}
\author[a]{Y. Shachar}
\author[b]{and S. Vaintraub}

% The "\note" macro will give a warning: "Ignoring empty anchor..."
% you can safely ignore it.

\address[a]{Weizmann Institute of Science, Rehovot, Israel}
\address[b]{Soreq Nuclear Research Center, Yavne, Israel}
\address[c]{Hebrew University of Jerusalem, Jerusalem, Israel}

\cortext[a]{Corresponding author.}

\begin{abstract}

A novel experiment has been commissioned at the Weizmann Institute of Science for the study of weak interactions via a high-precision measurement of the beta-neutrino angular correlation in the radioactive decay of short-lived $^{6}$He. The facility consists of a 14 MeV $d+t$ neutron generator to produce atomic $^{6}$He, followed by ionization and bunching in an electron beam ion source, and injection into an electrostatic ion beam trap. This ion trap has been designed for efficient detection of the decay products from trapped light ions. The storage time in the trap for different stable ions was found to be in the range of 0.6 to 1.2 s at the chamber pressure of $\sim$7$\times$10$^{-10}$ mbar. We present the initial test results of the facility, and also demonstrate an important upgrade of an existing method \cite{stora} for production of light radioactive atoms, viz.  $^{6}$He, for the precision measurement. The production rate of $^{6}$He atoms in the present setup has been estimated to be $\sim 1.45\times10^{-4}$ atoms per neutron, and the system efficiency was found to be 4.0$\pm$0.6\%. An improvement to this setup is also presented for the enhanced production and diffusion of radioactive atoms for future use.

\end{abstract}


\end{frontmatter}
%\linenumbers
%\linenumbers

\section{Introduction}

One of the possibilities to study fundamental interactions and the underlying symmetries is via precision measurements of the parameters of beta decay, viz. beta-neutrino angular correlation ($\alpha_{\beta\nu}$) of trapped radioactive atoms and ions, thus probing the minute experimental signal that originates from possible tensor or scalar terms in the weak interaction of beyond-the-standard-model nature \cite{severijns001, Oscar001}. For precision measurements of beta-neutrino correlation, ion traps are convenient tools since the recoiling nuclei or atom, subsequent to the beta decay, are at sub-keV energies, for which determination of their energy and momentum is very difficult without using the conditions in traps. The concept of the measurement is to produce $\beta$-emitter radioactive atoms ($^{6}$He in the present case), ionize and accelerate them to a few keV's, and then trap them inside an electrostatic ion beam trap (EIBT). The decay products from the trapped ions will be measured in coincidence to determine a valid correlation event. The recoiling daughter nuclei will be detected by a position-sensitive microchannel plate detector inside the trap and the electrons will be measured in a large area position-sensitive detector (LAPS) on the outside of the trap chamber. The working design of the LAPS detector has been published in Ref. \cite{laps}. Thus, $\alpha_{\beta\nu}$ can be reconstructed from the kinematics of the electron and recoiling daughter nucleus, which will be known to good precision.

The $^{6}$He nucleus is a good candidate for weak interaction studies since (a) it decays to the ground state of $^{6}$Li, except a very small branching ratio of (1.65$\pm$0.10)$\times$ 10$^{-6}$ for $d+\alpha$ channel \cite{6He-branching-ratio}, (b) the transition being $0^+ \rightarrow 1^+$ is a pure Gamow-Teller decay\footnote{The expected precision for the $^{6}$He experiment is below 1\%, so higher order corrections also must be considered. The detailed analysis of these corrections for the $^{6}$He $\beta-\nu$ correlation is found in \cite{Frank}} makes it independent of the nuclear matrix elements andthus free from nuclear model structure uncertainties, (c) it has a half-life of 806.89$\pm$0.11 ms \cite{6He-halflife1}, which is sufficient for ionization and injection into a trap where its decay can be measured, (d) it has a large Q value of $\sim$3.5 MeV to clearly detect its signature, and (e) helium, being a noble gas, does not react with residual gasses during extraction. Very few measurements have been performed for the beta-neutrino correlation for $^{6}$He \cite{6He-precision1,6He-precision2,6He-precision3} with high precision. Our scheme aims at reducing the systematic errors and achieving sub-percent precision in the measurement.

We have embarked on an experimental program to study the angular correlation $\alpha_{\beta\nu}$ from the decay of trapped, light, radioactive ions inside an EIBT \cite{hass_lightnuclei}. Electrostatic traps have been used extensively in atomic physics research and can be used to trap ions of a few keV's, irrespective of mass and charge of the species \cite{eibt-zafman}. The use of such a trap for fundamental interaction studies has not been previously reported and it exhibits potentially very significant advantages over other trapping schemes. In this report, we present the construction and results of test runs of the trapping of stable ions in the EIBT for measurement of storage time. We also describe the production of $^{6}$He neutral atoms necessary for feeding into the electron beam ion source for accumulation and ionization and then trapping in the EIBT for decay measurements. This scheme is not confined to $^{6}$He; it allows  the production of light radioactive ions and the study of their decay properties in smaller laboratories, which will be complementary to big collider-based physics programs.

In the next sections, we discuss the  main components of this project. The design and test results of ion trapping in the EIBT are discussed in Section \ref{sec:eibt}; the electron beam ion source used for ionizing and bunching is discussed in Section \ref{sec:ebis}, and Section \ref{sec:6he} presents the details of the experimental production of $^{6}$He atoms and related calculations for the production and transmission efficiency in the system. Envisaged improvements and future prospects of this work are discussed in Section \ref{sec:future}. 

\section{Electrostatic Ion Beam Trap}
\label{sec:eibt}

The electrostatic ion beam trap (EIBT) was developed at the Weizmann Institute of Science for trapping low-energy (few keV) ions in the time range of a few milliseconds to seconds \cite{eibt-zafman}. The storage time of such small table-top devices is limited by the residual gas in the chamber. The detailed principle of operation has already been presented in an earlier publication \cite{eibt-dahan}. The trap employed in our experiments employs a similar principle of operation but incorporates several features that were not part of the basic EIBT design. The current trap design is specially crafted to provide efficient detection of the $\beta$-decay products from the trapped light radioactive ions, such as $^{6}$He$^{+}$ and $^{16}$N$^{+}$. The length of the trap was chosen such that the detector placed inside the trap will have larger solid angle and higher collection efficiency from $^{6}$He decay products after considering the recoil energies. For other nuclei, the beam energy and the position of the detector can be changed to maximize the detection efficiency. 

\begin{figure}
\centering
\includegraphics[width=1.\linewidth]{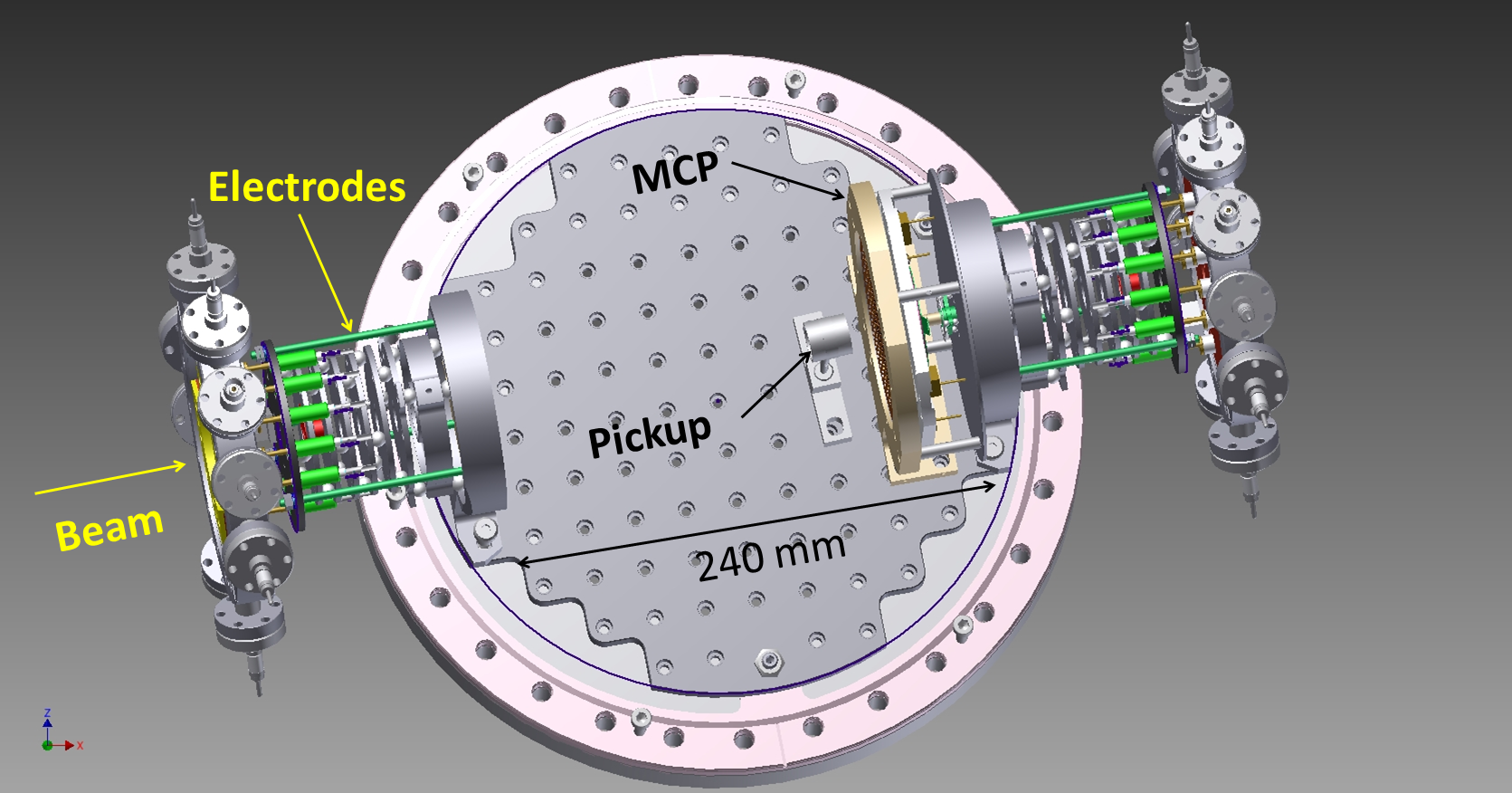}
\caption{Top view of 3D Inventor model of the Electrostatic Ion Beam Trap used for trapping low energy ions. A pickup used for diagnostic purposes and a position-sensitive microchannel plate detector for detection of neutral particles are visible in the drawing. }
\label{fig_TrapInv}
\end{figure}

The EIBT consists of two sets of eight electrodes, acting both as electrostatic mirrors and as lenses by producing a retarding field, which reflect the beam along its path and focus it transversely, respectively. The innermost electrodes are grounded to create a field-free region where the ion beam has a well-defined kinetic energy and direction in space. By applying a small RF voltage of a few volts on one of the mirror electrodes, the ion beam can be bunched. The entrance mirror is grounded upon injection of the ion bunch. Once the ion bunch fills the entire trap, the potentials on the entrance mirror are raised quickly ($\sim$50 ns) so that the ion bunch oscillates back and forth between the two mirrors. A capacitive pickup situated inside the innermost electrode of one of the mirrors is used to detect the traversing of the ion bunch. Another pickup was placed between the mirrors for beam diagnostic purposes. A 3D inner view of the EIBT displaying the electrostatic mirrors, a pickup and a large area microchannel plate with a resistive anode encoder (MCP-RAE) and 6.4 mm diameter central hole \cite{mcp-web} is shown in Fig. \ref{fig_TrapInv}. In order to minimize systematic errors, the mirrors and pickup are mounted on a precision optical table to ensure precise alignment.

\subsection{Trapping with stable ions}
The trapping system was tested with stable ions to measure storage times for different ion species. The inlet of the ion source (Sec. \ref{sec:ebis}) was connected to gas cylinders filled with CO and $^{4}$He gas. The gas flow was controlled by a motor-operated needle valve. 4.2 keV singly charged ions of different masses, viz. CO$^{+}$, O$^{+}$ and $^{4}$He$^{+}$, were used for testing the trap system. The source was operated in pulsed mode and an electrostatic beam chopper was used to further reduce the ion bunch size. The width of the chopped pulse was varied in different test runs to obtain the best suited ion beam bunch. The momentum selection of the ions was performed by a dipole bending magnet upstream of the trap. Moreover, the optimized voltages for the EIBT's mirrors were calculated using the SIMION \cite{simion-web} package to get the best trapping parameters, and subsequently were applied on the actual trap mirrors.  The base pressure in the EIBT was $\sim 7\times 10^{-10}$ mbar. \\

\begin{figure}[!t]
\centering
\includegraphics[width=1.\linewidth]{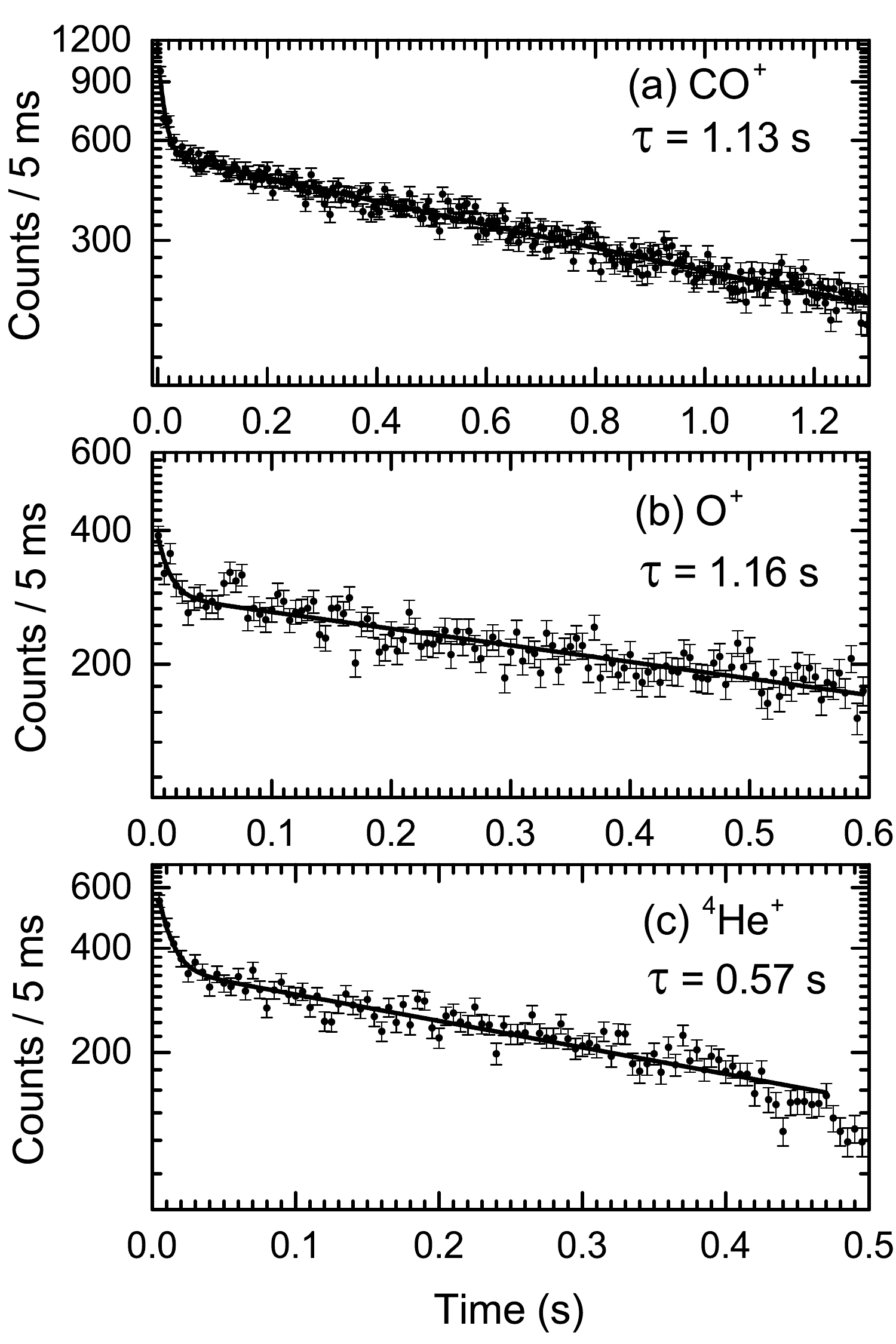}
\caption{Decay of different ions bunch trapped inside the electrostatic ion beam trap at the base pressure of $\sim 7\times 10^{-10}$ mbar. Data was fitted with double exponential decay (solid line). The first component corresponds to the loss of ions that entered the trap on unstable trajectories. The second component corresponds to the decay time of the ion bunch or the storage time.}
\label{fig_Trap}
\end{figure}

The positively charged ions can be scattered by collisions with the residual gas atoms during their oscillations between the  mirrors. Other ions undergo electron capture and are neutralized. Both of these processes produce losses from the trapped ion beam and depend largely on the pressure inside the trap and thus yield a pressure-dependent storage time. The decay curves for CO$^{+}$, O$^{+}$ and O$^{+}$ ions are shown in Fig. \ref{fig_Trap}(a), \ref{fig_Trap}(b) and \ref{fig_Trap}(c), respectively. The ion current was higher for CO$^{+}$ ions in comparison to O$^{+}$, and thus the statistics are better in the CO$^{+}$ case. The decay curves were fitted with a two-component exponential decay function. The first component is a fast decay associated with the loss of ions that entered the trap on unstable trajectories. This time is less than 10 milliseconds. The second component, corresponding to the decay time of the ion bunch at a given pressure, is termed the storage time of the beam inside the EIBT. We achieved storage times of $\sim$1.2 s for O$^{+}$ and CO$^{+}$ ions, and $\sim$0.6 s for $^{4}$He$^{+}$ ions. These storage times can be increased by improving the vacuum inside the chamber. However, $^{6}$He being heavier than $^{4}$He will have a longer storage time because it travels slower for the same energy. Thus, the currently-achievable pressure should give a storage time for $^{6}$He that is comparable to its half life.

\section{EBIS}
\label{sec:ebis}

\begin{figure}[!ht]
\centering
\includegraphics[width=1.\linewidth]{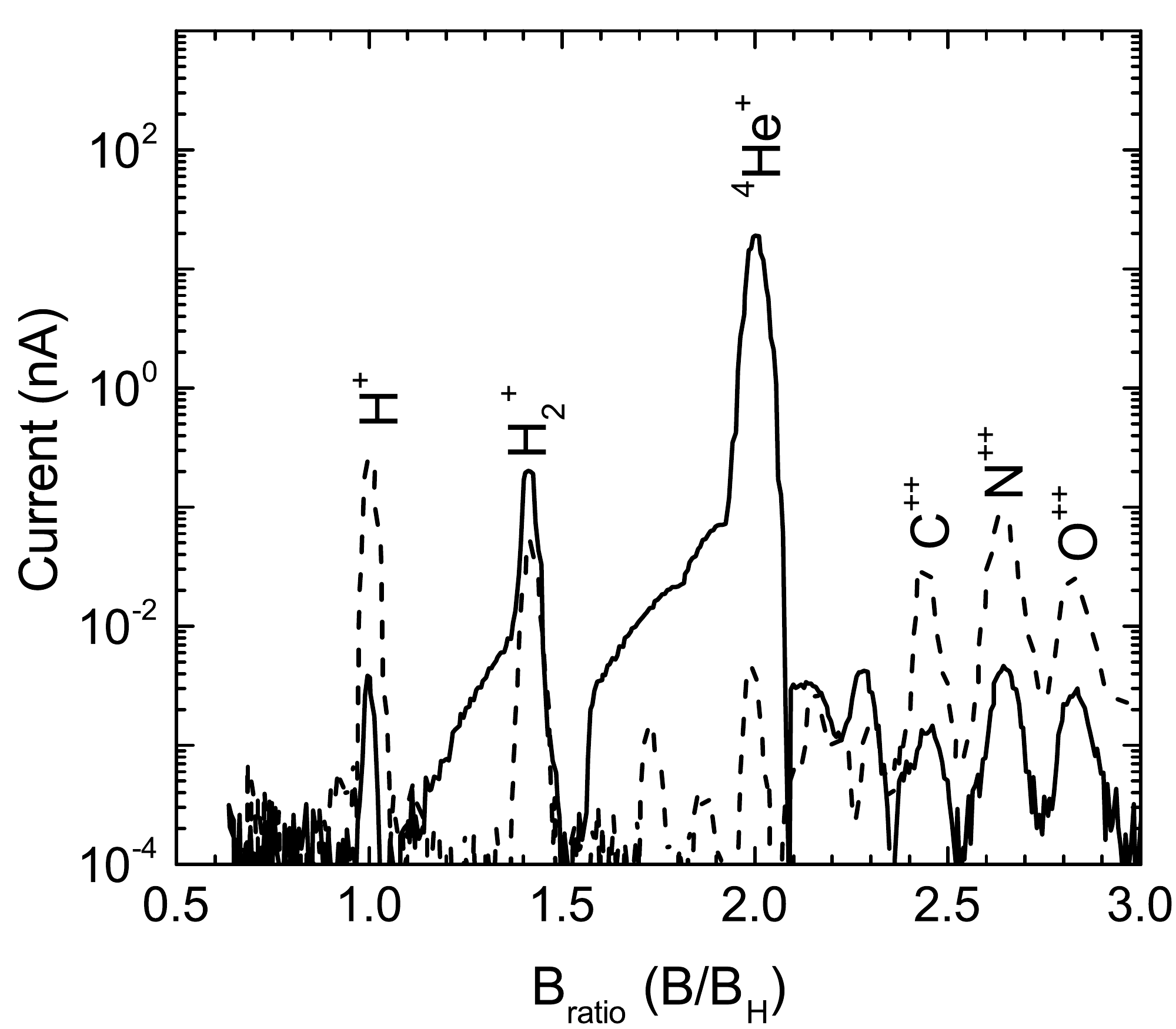}
\caption{Mass scan for different gas injected to the EBIS as analyzed by the 30$^{\circ}$ bending magnet. The ion source was operated in DC mode. Faraday cup current is plotted against the ratio of magnetic field to the magnet field for hydrogen. Input gases were helium (solid line) and methane (dash line). The $^{6}$He peak should appear at the same location as the C$^{2+}$ peak and it is clearly resolved from the peaks of neighboring masses. }
\label{fig_MagnetScan}
\end{figure}

The electron beam ion source (EBIS) was purchased from Dreebit GmbH, Germany \cite{dreebit-web}. The purpose of the EBIS is the ionization, accumulation and bunching, and injection of the radioactive and/or stable atoms into the trap. Additionally, the EBIS is meant to accumulate the tiny amount of produced radioactive isotopes before injection into the EIBT, thus increasing the overall efficiency. The EBIS has been modified for the requirements of this project to enhance trapping and efficient charge-breeding \cite{ebit}. The gas injection assembly of the ion source was modified because of the very low flux of $^{6}$He produced in our scheme. In the standard Dreebit EBIS, the gas is injected through a capillary tube outside the drift tube assembly. This results in some of the ions being pumped away before trapping. Another loss mechanism for $^{6}$He, even if there were no pumping, is that the atoms bounce around for some time inside the EBIS chamber until they `find' the drift tube. Some of the $^{6}$He ions decay in this time. In this EBIS, the drift tube design was modified to inject gas directly into the drift tube, which results in increased trapping and ionizing efficiency.  An EBIS can produce highly charged positive ion bunches in the few keV range, suitable for injection and trapping in the electrostatic trap. The time during which the atoms are confined inside the drift tube, the ionization time, is important  for getting higher charge states. The EBIS was tested for different ionization times, and a few tens of ms served as a good time for singly- or doubly-charged ions. The charge distribution produced by the EBIS and the momentum selection by the dipole magnet are presented in Fig. \ref{fig_MagnetScan}. The magnet has a good mass selection, $\Delta$M = 1, in our region of interest (M = 6). Higher masses were also clearly separated by this magnet, but are not shown in the graph. 

\section{Production of the $^{6}$He atoms}
\label{sec:6he}

\begin{figure}[!h]
\centering
\includegraphics[width=1.\linewidth]{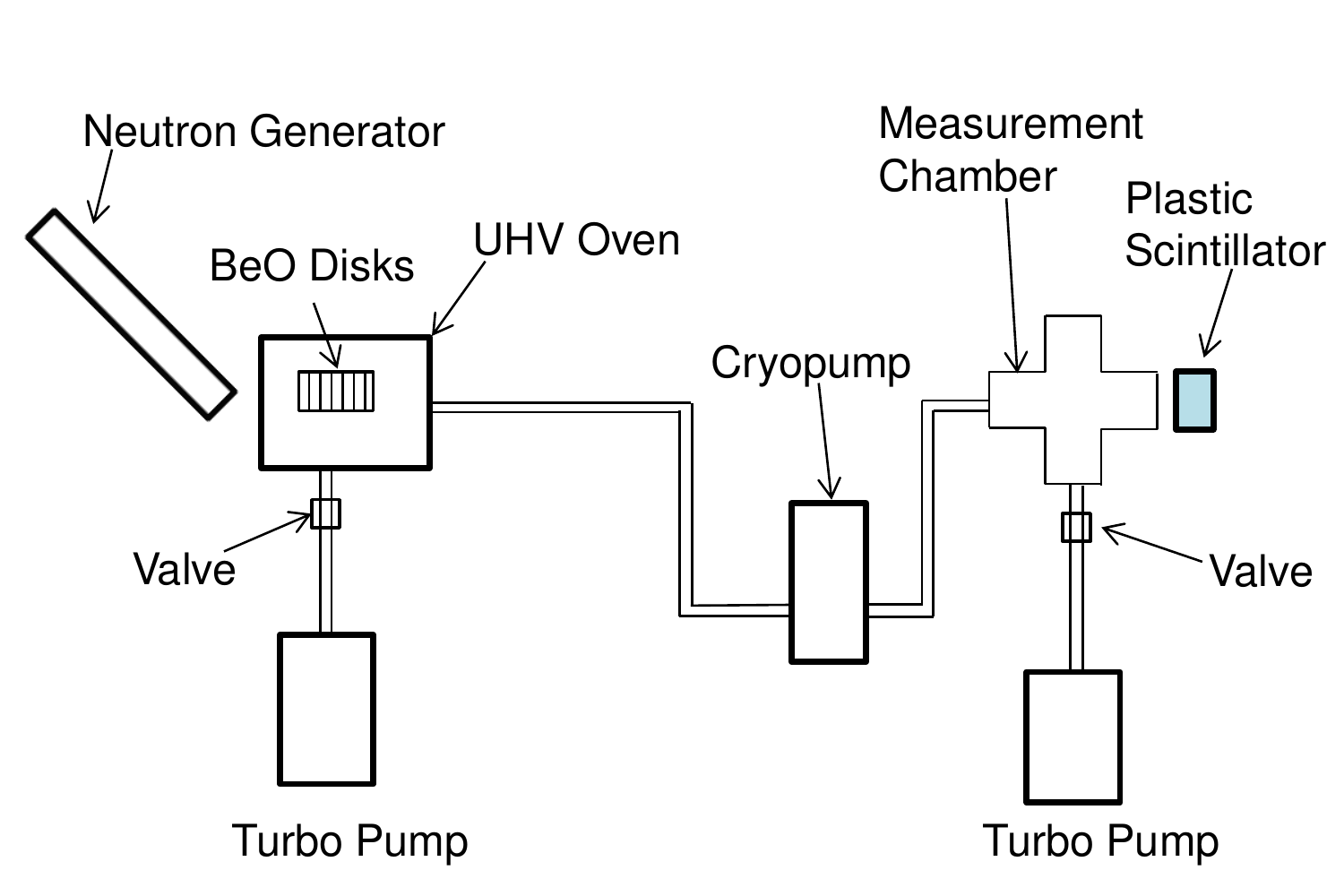}
\caption{Block diagram of the $^{6}$He production experiment.}
\label{fig_Block}
\end{figure}

$^{6}$He atoms were produced using the n($^{9}$Be,$\alpha$)$^{6}$He reaction. A block diagram of the production and measurement setup is shown in Fig. \ref{fig_Block}. Isotropic neutrons with an average kinetic energy of 14 MeV were produced using a commercial \textit{d+t} neutron generator (NG) manufactured by VNIIA \cite{vniia}. A UHV oven \cite{heatwave-web}  was used to heat a stack of 38 porous $\phi$ 38 mm $\times$ 2 mm BeO fibre disks to 1400$\pm$1$^\circ$C. This oven can be maintained at a vacuum of $\sim 10^{-8}$ mbar, and it can accommodate up to 80 disks. The area surrounding the oven was heavily shielded against neutron damage to detectors and neutron-activation contamination to the measurement. The effused gases from the oven passed through a cryopump to trap impurities. Filtered $^{6}$He atoms were finally measured inside the measurement chamber. The measurement chamber was a six-way cross made of stainless steel. The outer diameter of the tube was 63.5 mm and the edge-to-edge length of the chamber was 356.5 mm. The detection end of the chamber was fitted with 75 $\mu$m thick mylar window in order to minimize energy loss by electrons. The oven and the measurement chamber were connected to the cryopump by a tube with inner diameter of 3.8 mm. The length of tube from oven to the cryopump was $\sim$4 m and from cryopump to the measurement chamber was $\sim$5 m. There were pneumatic valves at the entrance and the pumping port of the measurement chamber that were controlled by a LabVIEW program. During the measurement, the valve between the measurement chamber and the turbopump was closed in order to prevent helium from escaping the chamber. The pressure in both the production and the measurement chamber was $\sim$10$^{-8}$ mbar.

\begin{figure}
\centering
\includegraphics[width=1.\linewidth]{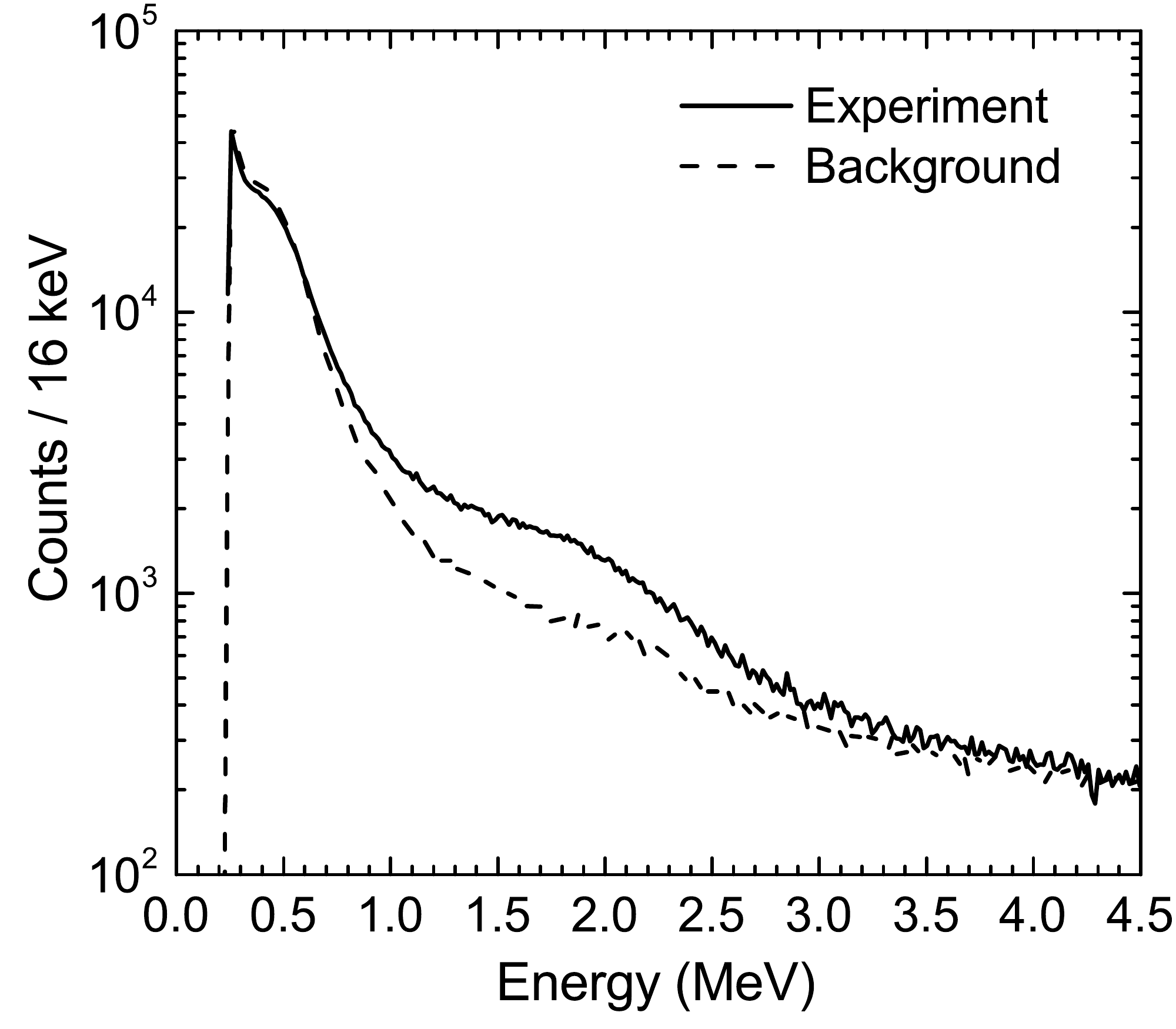}
\caption{The spectrum of $^{6}$He decay inside the measurement chamber detected by the plastic scintillator. The online measurement is denoted by the solid line whereas background for same time is denoted by the dash line.}
\label{fig_6HeExp}
\end{figure}

A plastic scintillator was used to measure the $\beta$ radiation from the decay of $^{6}$He. The detector was calibrated with a $^{90}$Sr $\beta$-emitter source that decays to $^{90}$Y, which in turn undergoes $\beta$ decay. The end point energy of 2.28 MeV was used for calibration. The $^{6}$He production measurement was performed in cycles of 12.6 second. Each cycle consisted of operation of the NG at 30 Hz for 2.5 s, followed by a rest time of 100 ms, and then acquiring signals from the plastic detector for a period of 10 s. A total of 900 cycles were executed. An Ortec model 919E ADC, in combination with Ortec Maestro software, was used for data acquisition. A LabVIEW program provided the gate pulse for the operation of the NG and data acquisition. The background was also recorded under the same conditions.  The experimental spectra detected by the plastic scintillator are shown in Fig. \ref{fig_6HeExp}. The subtracted spectrum and its comparison with simulations are described in the next section and presented in Fig. \ref{fig_6HeSimExp}.

\subsection{Production and efficiency calculations}

Production and efficiency calculations could be performed using different Monte Carlo simulation techniques. In our calculations, the GEANT4 simulation package \cite{geant4-1,geant4-2} has been used extensively. The measurement chamber was simulated using the exact chamber dimensions with all flanges and mylar window. The physics list in the simulation consists of particle constructors, standard electromagnetic physics, and radioactive decays. $^{6}$He atoms were used as the primary ion source. The particles were generated randomly inside the measurement chamber walls. The majority of the particles hitting the plastic detector originated on the axis of the chamber which was coaxial to the detector axis. The plastic scintillator was used as the active detector material, and the  chamber walls were also made sensitive in order to get the detailed scattering information. 

\begin{figure}
\centering
\includegraphics[width=1.\linewidth]{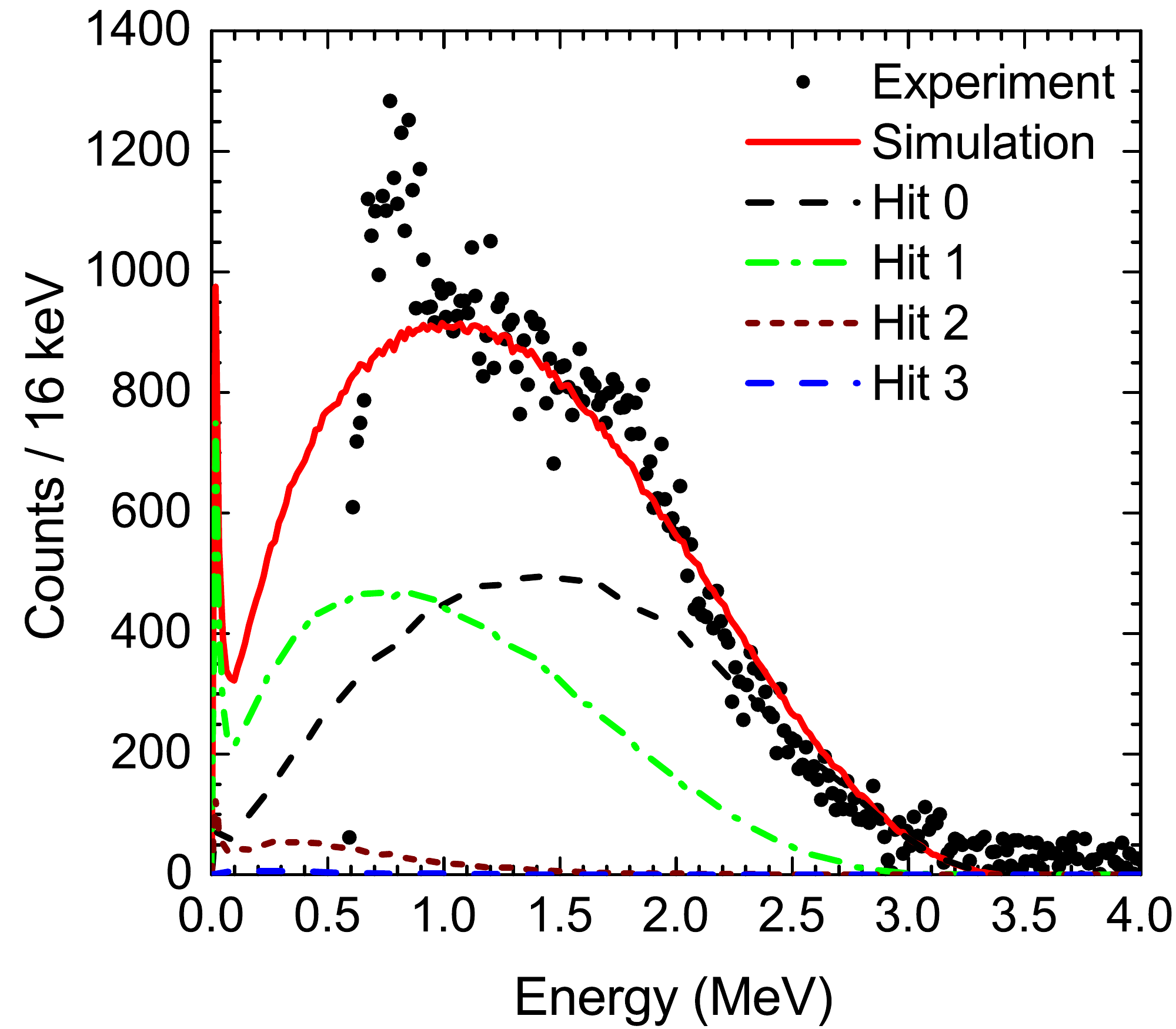}
\caption{(color online) Experimental and the GEANT4-simulated spectrum of decay of $^{6}$He atoms inside the measurement chamber. The background-subtracted experimental spectrum is denoted by black dots. The normalized total simulated response is denoted by the solid red line. Other lines correspond to the number of hits with the chamber walls followed by detection in the plastic detector.}
\label{fig_6HeSimExp}
\end{figure}

A total of $10^{8}$ decay events were simulated and the data file was recorded in CERN root \cite{root-1,root-2} format, which could be analyzed offline with different gating conditions. The data was analyzed to find the scattering of betas from the the chamber walls and then detection by the plastic scintillator, and the results are shown in Fig. \ref{fig_6HeSimExp}. The simulated spectrum was  normalized to the experimental spectrum, and the normalization factor was found to be 0.040$\pm$0.002. This implies that (4.0$\pm$0.2) $\times10^6$ atoms reached and decayed inside the measurement chamber.

\begin{figure}
\centering
\includegraphics[width=1.\linewidth]{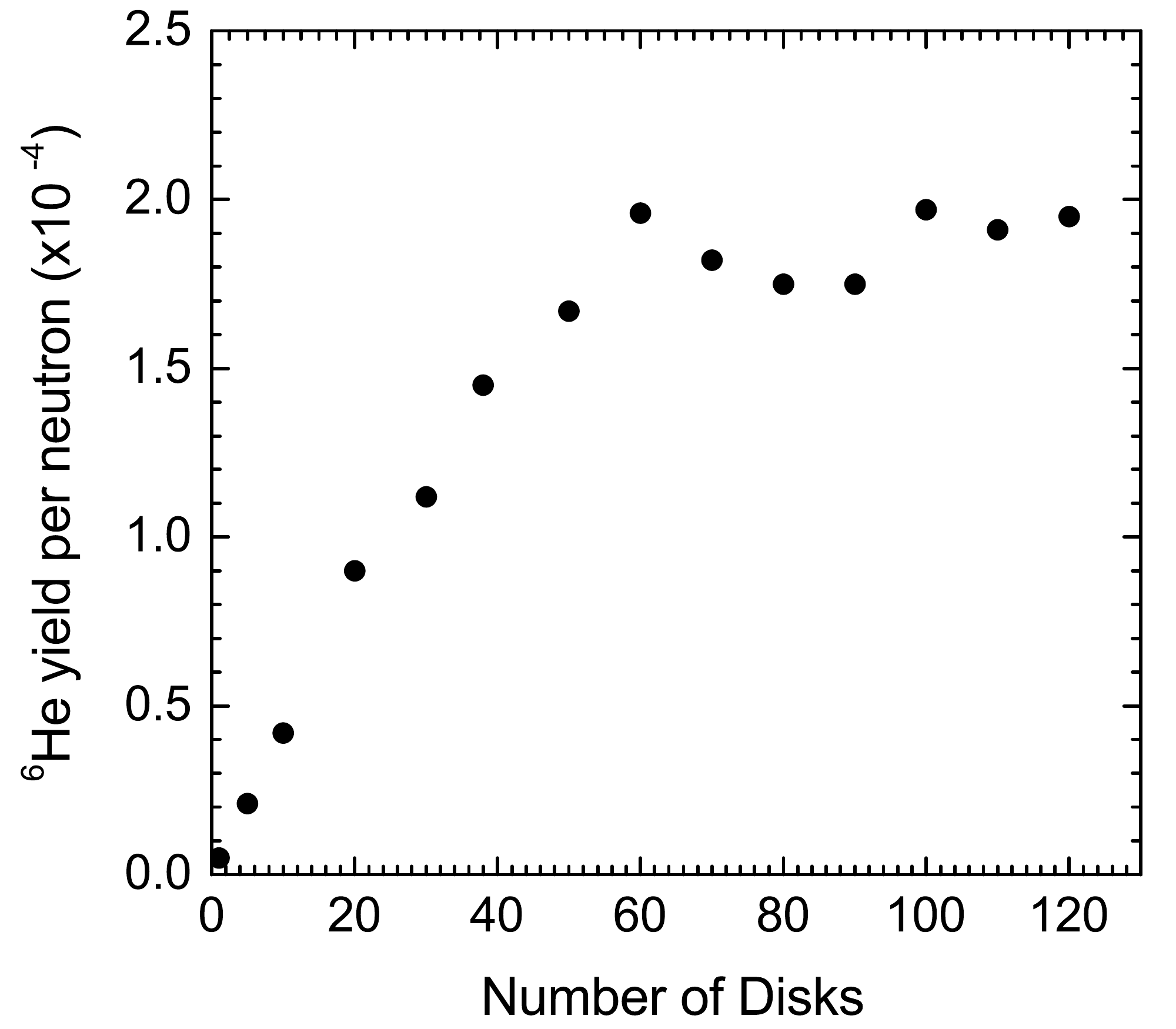}
\caption{GEANT4 simulation of production yield of $^{6}$He as a function of number of BeO disks.}
\label{fig_6HeProductionSim}
\end{figure}

The production yield of $^{6}$He atoms in the current geometry was also calculated using a GEANT4 simulation that incorporated the full production profile in the thick target, and also included the $^{9}$Be(n,2n)$^{8}$Be reaction that shifts the neutron energy spectrum of the generator towards lower energies. The temperature dependence of the diffusion of $^{6}$He atoms from these disks was not incorporated in this simulation. The release fraction of $^{6}$He in a similar reaction, as a function of temperature, has been measured by Stora et al. \cite{stora}. However, the dimensions of the BeO disks in our case are larger, and thus we expect at least 82\% release of $^{6}$He in our experiment. Each simulated run consisted of 10$^{6}$ events from an isotropic point source of neutrons for a given number of BeO disks. The production yield of $^6$He atoms per neutron as a function of the number of BeO disks is shown in Fig. \ref{fig_6HeProductionSim}. 38 disks were used in our experiment for which the production yield is $1.45 \times 10^{-4}$ atoms per neutron. The neutron flux from the commercial neutron generator (NG) was measured to be 2.2 $\times 10^{8}$ neutrons/pulse with an uncertainty of $\pm$10\%. The ratio of $^{6}$He atoms decaying inside the measurement chamber to the atoms produced, the measured efficiency ($\epsilon_{\text{measured}}$), was found to be 0.19$\pm$0.03\%.

The calculation of the system efficiency, i.e. the fration of the number of $^{6}$He atoms reaching the measurement chamber to the number of $^{6}$He atoms produced, requires a few corrections to be applied to the measuread efficiency and is given by $\epsilon_{\text{system}} = C_{\text{D}} \cdot C_{\text{V}} \cdot \epsilon_{\text{measured}}$, where $C_{\text{D}}$ is the decay correction and $C_{\text{V}}$ is the volume correction. The first correction ($C_{\text{D}}$) is due to the decay of $^{6}$He during the production and the diffusion. Any atom decaying during diffusion will not result in a signal in the plastic detector due to the absorption of beta particles in the thick walls. $C_{\text{D}}$ is dependent on the half-life, the duration of irradiation, and the measurement time \cite{Filippone}. For $^{6}$He, we found this factor to be 2.7. The second correction $C_{\text{V}}$ is the ratio of the total volume to the measurement chamber volume. $^{6}$He diffusing from the oven to the measurement chamber will fill the volume of the tubes and cryopump, which is significant enough to affect the system efficiency. The factor $C_{\text{V}}$ was found to be 8.1. After correction for these factors and taking into account the uncertainties in the neutron flux and simulations, we found a system efficiency of 4.0$\pm$0.6\%.

\begin{figure}
\centering
\includegraphics[width=1.\linewidth]{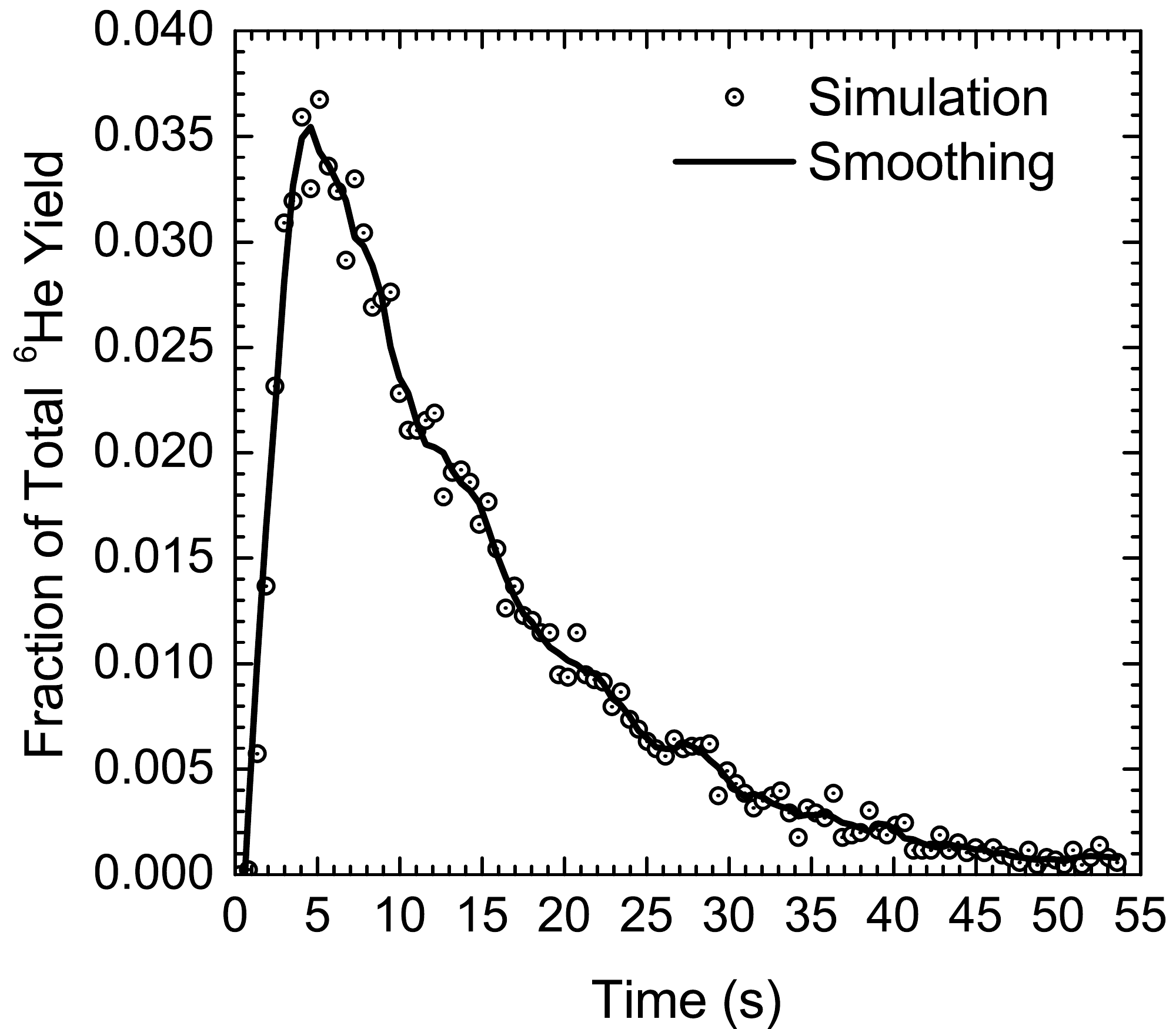}
\caption{Simulation for the fraction of $^{6}$He atoms reaching the measurement chamber as a function of time after leaving the oven.}
\label{fig_6He_transmission}
\end{figure}

A Monte Carlo simulation was also performed to get an estimate for the system efficiency. This simulation takes into account the transit time of atoms through the connecting pipes and the cryopump between the oven and the measurement chamber, and it also considers the decay during travel. In the simulation, around 8000 atoms were fired through the pipe. The simulated response, i.e., the fraction of $^{6}$He atoms reaching the measurement chamber as a function of time are shown in Fig. \ref{fig_6He_transmission}. The average transit time was found to be slightly more than 10 s. Thus a large fraction of the  atoms decayed during transit to the measurement chamber. According to the full simulation, the system efficiency that includes the diffusion time and the in-travel decay loss comes out to be 3.8\%. The simulated result agrees well with the experimental value of 4.0$\pm$0.6\% quoted above.  

It is important to note that the results obtained with the present scheme demonstrate that virtually all $^{6}$He atoms produced by the \textit{n}-induced reaction leave the target assembly in a short time compared to the half-life of $^{6}$He. As such, it is an important validation of the large scale up of the present target mass of BeO compared to the BeO target in the Isolde experiment \cite{stora}. This observation bodes well for possibly even larger scale-ups in future experiments.

\section{Future prospects}
\label{sec:future}

There are three main areas of improvement for the present scheme: first, an increase in the neutron yield which will lead to a higher production rate of radioactive atoms; second, the reduction in transit loss of radioactive atoms; and third, an increase of the storage time of ions inside the EIBT. 

The use of the neutron generator will be replaced and upgraded to the accelerator-based neutron source at the Soreq Applied Research Facility (SARAF) \cite{saraf1} at Soreq Nuclear Research Center, Israel. SARAF houses a high yield neutron generation facility using medium energy deuterons (5 MeV) on the  liquid lithium target (LiLiT) \cite{LiLiT}. With the expected current of deuterons of 1 mA and an increase in the number of BeO disks to 80, the expected yield of $^{6}$He production at SARAF-I is expected to be at least two orders of magnitude higher than in the current experiment. The second stage of SARAF (SARAF-II) is expected to provide another 2-3 orders of magnitude increase in the neutron intensity. The transit loss of the neutral atoms is planned to be reduced manyfold using larger diameter (4 inch) connecting pipes. Such pipes have already been installed at SARAF for future experiments and test runs are ongoing. The larger pipe will reduce the average transport time from 10 seconds to slightly below 1 second. This will result in improvement of transmission efficiency to nearly 90\%. The storage time in the EIBT is inversely proportional to the residual pressure. We are working on reducing the pressure by an order of magnitude by changing the vacuum pump configuration and by baking the chamber. With a higher neutron yield, better transmission efficiency and increased storage time, this experiment provides a promising avenue for measuring $\alpha_{\beta\nu}$ with higher precision and less systematic uncertainty in a short period of time. 

\section{Summary}
\label{sec:summary}

We presented a new experimental scheme for the search of physics beyond the standard model based on trapping low-energy radioactive ion beams ($^{6}$He in the present case) in an electrostatic ion trap for  measuring the beta-neutrino  angular correlation ($\alpha_{\beta\nu}$) from their $\beta$-decay. This is the first use of an electrostatic ion beam trap in such studies and provides a complementary solution to big accelerator-based research programs. This kind of experiment can be performed in smaller laboratories. The results for production of $^{6}$He atoms and their diffusion were discussed. The $^{6}$He production yield, in the current configuration, has been estimated to be $1.45 \times 10^{-4}$ atoms/neutron, and the system efficiency was found to be 4.0$\pm$0.6\%. The EBIS has been used for ionizing, bunching and accelerating the ions prior to injection into the EIBT where they are stored for a few seconds for decay measurements. The storage time  was found to be sufficient to efficiently detect the decay of radioactive $^{6}$He ions. The experimental system can open new avenues in the field of precision measurements at a high neutron flux facility, e.g., at SARAF, by producing and trapping different light radioactive ions at  higher yields.  

\section*{Acknowledgments}
The authors would like to acknowledge the support from the Israel Science Foundation (ISF), the European Research Council (ERC), the PAZY Foundation and the DOE-IAEA cooperation fund.

\bibliographystyle{elsarticle-num}
\bibliography{article}


%% References

\end{document}